\pdfoutput=1
\documentclass[11pt]{article}

\usepackage[margin=1in]{geometry}
\usepackage{amsmath, amssymb, amsfonts}
\usepackage{graphicx}
\usepackage{hyperref}
\usepackage{booktabs}
\usepackage{enumitem}
\usepackage{natbib}
\usepackage{amsthm}
\newtheorem{definition}{Definition}
\usepackage{microtype}
\usepackage{float}
\usepackage{caption}
\usepackage{subcaption}
\usepackage{mathtools}
\usepackage{xcolor}
\usepackage{tikz}
\usepackage{placeins}
\usetikzlibrary{positioning,arrows.meta,calc,shapes.geometric}

\definecolor{gaBlue}{HTML}{2563EB}
\definecolor{gaBlueDark}{HTML}{1E40AF}
\definecolor{gaCoral}{HTML}{DC2626}
\definecolor{gaCoralDark}{HTML}{991B1B}
\definecolor{gaPurple}{HTML}{7C3AED}
\definecolor{gaPurpleLight}{HTML}{EDE9FE}
\definecolor{gaTeal}{HTML}{0D9488}
\definecolor{gaTealLight}{HTML}{CCFBF1}
\definecolor{gaAmber}{HTML}{D97706}
\definecolor{gaAmberLight}{HTML}{FEF3C7}
\definecolor{gaGreen}{HTML}{059669}
\definecolor{gaGreenLight}{HTML}{D1FAE5}
\definecolor{gaCoralLight}{HTML}{FEE2E2}
\definecolor{gaGray}{HTML}{4B5563}
\definecolor{gaGrayLight}{HTML}{F3F4F6}

\hypersetup{
  colorlinks=true,
  linkcolor=gaBlueDark,
  citecolor=gaTeal,
  urlcolor=gaPurple
}

\title{\textbf{Mixup Barcodes for Topology-Aware Financial Decision Making}}
\author{
Buddha Nath Sharma$^{1}$, Joe Opitz$^{2}$, Adam Moser$^{2}$, \\
Sayam Palrecha$^{3}$, Sushovan Majhi$^{3}$, Atish Mitra$^{2}$ \\[0.3cm]
$^{1}$National Institute of Technology Sikkim, India \\
$^{2}$Montana Technological University, USA \\
$^{3}$George Washington University, USA
}
\date{}

\begin{document}

\maketitle

\begin{abstract}
We present a topology-aware system based on \emph{mixup barcodes} and persistent homology for financial decision making. The suggested approach uses topological summaries obtained from Takens delay embeddings of a univariate price series to quantify structural changes between reference and current market regimes. The $1$-Wasserstein distance between persistence diagrams, a mixup barcode disruption index, and persistence entropy divergence are combined to provide a novel stress score. A Golden Cross trading method is dynamically modulated by this score, which transforms a binary buy/sell signal into a continuous position-sizing process. At embedding settings chosen by maximizing in-sample Sharpe over a 72-point grid, in-sample assessment on the S\&P 500 and Bitcoin yields Sharpe ratios of $1.116$ and $1.238$, respectively, with much lower maximum drawdown compared with both Buy-and-Hold and the standard Golden Cross baseline. Our findings imply that topological summaries of market geometry include useful information that goes beyond traditional trend markers.

\end{abstract}

\section{Introduction}
\label{sec:introduction}

For a variety of reasons, including asset prices resulting from the interaction of numerous agents, macroeconomic pressures, policy interventions, and collective behavior~\citep{mantegna2000introduction,arthur2018economy}, financial markets are among the most complex dynamical systems investigated in quantitative science. This results in high-dimensional, nonlinear, and non-stationary time series~\citep{sornette2003critical,cont2001empirical}. The assumptions of stationarity and linearity that underpin traditional tools like autoregressive models, GARCH-family volatility estimators, and cointegration tests break down precisely during regime transitions and extreme events that are most relevant to risk managers and portfolio allocators~\citep{hamilton1989new}. This has led to a search for techniques that are sensitive to market data's \emph{geometric} structure rather than just its low-order statistics.

Such an approach is provided by \emph{Topological Data Analysis} (TDA) \citep{Carlsson2009TopologyAD, Dey_Wang_2022}. Its main tool, \emph{persistent homology}, summarizes topological features (like loops, voids, and connected components) in a \emph{persistence diagram} that is provably stable under perturbation of the input \citep{cohen2005stability} by tracking their birth and death across a one parameter filtration of simplicial complexes constructed from the data. A typical pipeline for time-series data moves a window over the series, uses the Takens delay embedding \citep{takens2006detecting} to lift each window into a Euclidean point cloud, and calculates the cloud's persistent homology as the window moves forward. This describes the geometry of the reconstructed attractor for a single series, which varies in shape depending on the local dynamical regime.

Persistent homology has primarily been used for regime discovery and crash early warning in quantitative finance. Gidea and Katz demonstrated that the $L^p$-norms of persistence landscapes grow strongly ahead of major crashes, such as the dot-com and 2007/2008 crises \citep{gidea2018topological}. Similar pre-crash elevation has been reported across ETF sectors and the 2020 COVID selloff \citep{aguilar2020topology}, across international markets \citep{guo2020empirical}, and through Betti-number dynamics in Asian indices \citep{yen2021using}. PH-based measures have also been used to monitor fragility and systemic risk in the Indian stock markets \citep{kulkarni2023investigation}. Similarly, PH-based metrics were utilized to investigate Granger-causal relationships between various financial instruments during crash periods~\citep{sharma2025causality} and identify crashes as Extreme Events in the stock market~\citep{rai2024identifying}; crash dynamics in cryptocurrency markets have been examined through complex-network analysis~\citep{mukhia2024complex}, and Euler-characteristic summaries have been used for interpretable classification of time series~\citep{sharma2026interpretable}. In addition to crash detection, topological features have been applied to machine-learning of Bitcoin transitions \citep{gidea2020topological}, regime clustering \citep{majumdar2020clustering}, and improved index tracking \citep{goel2020topological}. \citet{souto2023topological} discovered that the Wasserstein distance between persistence diagrams is a statistically significant predictor of realized tail risk. \cite{song2025can} looked into the possibility of using topological transitions in cryptocurrency systems as early warning indicators for severe swings in conventional markets. \cite{guritanu2025topological} created an interpretable framework that gives early warning signs for financial crises by extracting PH-based topological signals from multivariate return time series.

These frameworks are characterized by a common limitation: typical topological summaries depict a \emph{single} point cloud in isolation. They measure the complexity or uniqueness of the existing market geometry, but they do not quantify the disruption of that geometry in connection to a particular reference regime. This is a relational and directed inquiry that is much closer to what a risk manager actually asks.

Quantitative trading has historically relied on technical trend-following rules, among which the \emph{Golden Cross} remains a prominent technique. The mechanics of this strategy are fundamentally binary. It dictates establishing a long position when the 50-day simple moving average (SMA) crosses above the 200-day SMA---signaling an anticipated uptrend---and liquidating upon the inverse \emph{Death Cross}, wherein the short-term average falls below the long-term average. By leveraging the secular upward drift of equity markets, this crossover aims to maintain market exposure during bull regimes while shifting to cash during protracted downturns, thereby tolerating minor whipsaw effects to mitigate severe bear-market drawdowns. Historically, this approach has reduced the maximum drawdown of the passive buy-and-hold strategy by approximately 50\% on the S\&P 500, at the cost of only a marginal reduction in absolute returns. Furthermore, it is characterized by low trading frequency, generating merely a few dozen signals over several decades with average holding periods of roughly one year.

The academic literature provides mixed evidence regarding the efficacy of such technical rules. \cite{brock1992simple} demonstrate that moving-average and trading-range-breakout rules significantly outperform passive benchmarks on the Dow Jones Index, yielding returns inconsistent with random-walk and GARCH null hypotheses. Similarly, \cite{moskowitz2012time} document robust time-series momentum across 58 futures markets, a finding extended over a century-long horizon by \cite{hurst2017century}. Conversely, the predictive edge of these heuristics is tenuous, and their mechanical formulations are inherently coarse. After correcting for data-snooping bias across a comprehensive universe of technical rules, \citet{sullivan1999data} reveal that out-of-sample profitability largely dissipates, and \citet{huang2020testing} find that moving-average rules implemented on ETFs deliver lower average returns and Sharpe ratios than buy-and-hold.

More fundamentally, the Golden Cross provides strictly \emph{binary} exposure---fully invested or entirely in cash---lacking any mechanism for dynamic risk scaling. As a lagging indicator derived from smoothed prices, it results in delayed entries and exits, suffers from repeated whipsaws in range-bound markets, and exhibits inadequate responsiveness to rapid market dislocations, such as the March 2020 crash \citep{rai2022sentiment}. The passive buy-and-hold alternative, while avoiding transaction-induced friction, remains fully exposed to the maximum drawdown of every systemic crisis (exceeding $30\%$ for equities and $80\%$ for Bitcoin in our sample), rendering it suboptimal for risk-constrained portfolio allocations. Crucially, both paradigms condition exclusively on absolute price levels and low-order trend statistics, thereby discarding the \emph{geometric} structure of local price dynamics, which encapsulates the earliest signatures of regime transitions.

These two limitations are mutually reinforcing. Traditional trend-following rules capture directional movement but ignore the underlying geometric structure of the market regime. Conversely, current topological indicators capture geometry but only in isolation, characterizing the present market state without measuring its deviation from a defined baseline. We address both gaps simultaneously using a unified framework: the \emph{mixup barcode} introduced by \citet{wagner2024mixup}. This metric tracks the degree to which the persistent topological features of a reference point cloud are compressed when a second point cloud is merged into the same space. Previously, this technique has been exclusively applied to learned embedding spaces \citep{wagner2024mixup,yadav2025missing}; a closely related Euler-characteristic analogue, the mixup Euler characteristic profile, has recently been used to detect regime transitions in dynamical systems \citep{majhi2026regime}, and stability theory for such interaction descriptors has begun to appear \citep{kawamura2026intersection}. By designating a calm baseline period as the reference and a moving window as the current state, we derive a normalized, $[0,1]$-valued metric of topological disruption. This measure is inherently relational and serves as an intuitive risk dial. Our primary contributions are summarized as follows:

\begin{itemize}
  \item To our knowledge, we provide the first financial application of the mixup barcode. Specifically, the mixup percentage $MP_1$ measures the extent to which the current market environment disrupts the loop topology of a baseline regime, offering a relational assessment rather than an isolated description of either state.

  \item We integrate $MP_1$ into a composite stress indicator, $S(t)$, alongside the $1$-Wasserstein distance between persistence diagrams and a divergence term based on persistence entropy.

  \item We leverage $S(t)$ to transform the traditional binary Golden Cross signal into a continuous, risk-aware allocation mechanism. In this framework, the trend filter dictates the direction of market exposure, while the topological measure determines its magnitude.

  \item Evaluated on the S\&P\,500 and Bitcoin, net of transaction costs and at the embedding parameters that maximize the in-sample Sharpe ratio, the resulting strategy attains Sharpe ratios of $1.116$ and $1.238$, respectively, while at least halving the maximum drawdown relative to both Buy-and-Hold and the conventional Golden Cross.
\end{itemize}

The remainder of this paper is structured as follows. Section~\ref{sec:background} reviews the foundational concepts of persistent homology and the Takens embedding. Section~\ref{sec:mixup} formalizes the mixup barcode and the mixup percentage, culminating in the construction of our composite stress score. Section~\ref{sec:strategy} details the proposed position-sizing strategy. Section~\ref{sec:results} presents the empirical performance evaluated on the S\&P\,500 and Bitcoin markets. Finally, Sections~\ref{sec:discussion} and~\ref{sec:conclusion} explore the broader implications and limitations of our approach before offering concluding remarks, while Appendix~\ref{sec:appendix} provides the comprehensive grid search over embedding parameters.

\section{Preliminaries}
\label{sec:background}

\subsection{Persistent Homology}

Consider a finite point cloud $X=\{x_1,\ldots,x_n\}\subset\mathbb{R}^d$. For $\epsilon\geq 0$, the \emph{Vietoris--Rips complex} $\mathrm{VR}(X,\epsilon)$ is defined as the simplicial complex in which each $k$-simplex corresponds to a $(k+1)$-element subset of $X$ with diameter no greater than $\epsilon$. As $\epsilon$ grows, these complexes form a nested filtration
\[
\mathrm{VR}(X,0)\subseteq\mathrm{VR}(X,\epsilon_1)\subseteq\cdots\subseteq\mathrm{VR}(X,\infty).
\]
Applying the $k$-th homology functor $H_k(\cdot;\mathbb{F})$ and tracking the maps induced by inclusion yields a \emph{persistence module}, which, by the structure theorem \citep{zomorodian2005topology}, admits a decomposition into interval modules. Each interval represents a topological feature that appears (is born) at scale $b$ and disappears (dies) at some $d>b$. The collection of such birth--death pairs $(b,d)$ constitutes the \emph{persistence diagram} $\mathrm{Dgm}_k(X)$, or equivalently the \emph{barcode} $\mathcal{B}_k(X)$ of half-open intervals $[b,d)$.
Features with large persistence $d-b$ reflect genuine geometric structure, whereas those with $d-b\approx 0$ are typically attributed to noise.
To facilitate statistical operations such as averaging, functional representations like the \emph{persistence landscape} \citep{JMLR:v16:bubenik15a} embed diagrams into a Hilbert space.

\paragraph{Stability.}
The stability theorem \citep{cohen2005stability,chazal2014persistence} guarantees that the bottleneck distance $d_\mathrm{B}$ between diagrams is bounded above by the Hausdorff distance $d_\mathrm{H}$ of the underlying point clouds,
\[
d_\mathrm{B}\!\left(\mathrm{Dgm}_k(X),\,\mathrm{Dgm}_k(Y)\right)\;\leq\;2\,d_\mathrm{H}(X,Y).
\]
Consequently, minor perturbations in the input data---whether from measurement noise or small market fluctuations---induce only bounded changes in the resulting diagram, thereby ensuring that topological signatures are inherently robust.

\subsection{Takens Delay Embedding}

Given a univariate time series $\{s_i\}_{i=1}^N$, the \emph{Takens delay embedding} with lag $\tau$ and embedding dimension $d$ constructs the point cloud
\[
P=\bigl\{(s_i,\,s_{i+\tau},\,\ldots,\,s_{i+(d-1)\tau})\bigr\}_{i=1}^{N-(d-1)\tau}\subset\mathbb{R}^d.
\]
Under mild genericity assumptions, Takens' theorem \citep{takens2006detecting} ensures that $P$ preserves the topological structure of the underlying dynamical attractor. In the present context, this embedding transforms a rolling window of normalized log prices into a point cloud whose geometric configuration captures the prevailing local market regime.

\subsection{Wasserstein Distance and Persistence Entropy}

The $p$-Wasserstein distance between two persistence diagrams $D_1$ and $D_2$ is defined as \citep{chazal2021introduction}
\[
W_p(D_1,D_2)=\left(\inf_{\eta:D_1\to D_2}\sum_{x\in D_1}\|x-\eta(x)\|_2^p\right)^{1/p},
\]
where the infimum is taken over all bijections $\eta$ that are permitted to map unmatched points to the diagonal. In this work we employ $W_1$, which measures the optimal matching cost under the Euclidean ground metric and is sensitive to both the count and the magnitude of topological differences between two market states. The \emph{persistence entropy} \citep{rucco2016characterisation} of a diagram $D$ quantifies the dispersion of feature lifetimes and is given by
\[
H(D)=-\sum_{(b,d)\in D} p_{(b,d)}\log p_{(b,d)},
\qquad p_{(b,d)}=\frac{d-b}{\sum_{(b',d')\in D}(d'-b')}.
\]

\section{Mixup Barcode}
\label{sec:mixup}

\subsection{Definition}

Let $A,B\subset\mathbb{R}^d$ denote finite point clouds corresponding to a \emph{reference} and a \emph{current} market state, respectively. Computing the persistent homology of $A$ in isolation produces birth--death pairs $(b,d)$; however, computing persistent homology on the union $A\cup B$ may cause features originally present in $A$ to terminate prematurely at some $d'\leq d$, since points from $B$ can fill in or merge cycles that were otherwise long-lived in $A$ alone.

\begin{definition}[Mixup Barcode]
Let $\mathcal{B}_k(A)=\{(b_i,d_i)\}$ denote the barcode of $A$, and let $\{(b_i,d'_i)\}$ denote the barcode of the image of the map $H_k(\mathrm{VR}(A,\epsilon))\to H_k(\mathrm{VR}(A\cup B,\epsilon))$ induced by the inclusion $A\hookrightarrow A\cup B$ (image persistent homology). Each bar of the image is born at the same scale $b_i$ as a bar of $\mathcal{B}_k(A)$ and dies at some $d'_i$ with $b_i\leq d'_i\leq d_i$, which matches the two barcodes one-to-one.
The \emph{mixup barcode} of $B$ acting on $A$ is defined as~\citep{wagner2024mixup}
\[
\mathcal{M}_k(A,B)=\bigl\{(b_i,\,d'_i,\,d_i):(b_i,d_i)\in\mathcal{B}_k(A)\bigr\},
\]
which records, for each feature, its birth $b_i$, its modified death time $d'_i$ in the union, and its original death time $d_i$ from $A$ alone.
\end{definition}

\subsection{Mixup Percentage}

Restricting attention to features with strictly positive persistence, $B_k=\{i:d_i-b_i>\epsilon_0\}$ with $\epsilon_0=10^{-10}$ (in practice, this threshold eliminates only the exactly diagonal bars arising from the reduction algorithm and imposes no near-diagonal filtering), the \emph{$k$-th mixup percentage} is defined as
\[
MP_k(A,B)=\frac{1}{|B_k|}\sum_{i\in B_k}\frac{d_i-d'_i}{d_i-b_i}\in[0,1].
\]
Each summand represents the fraction of its original lifetime that feature $i$ loses in the presence of $B$; the average thus captures the overall degree of topological disruption.
$MP_k=0$ indicates that $B$ leaves the $k$-dimensional topology of $A$ entirely undisturbed, corresponding to a regime that is topologically identical to the reference. $MP_k\approx 1$ signifies that nearly every $k$-dimensional persistent feature is annihilated almost immediately, reflecting a maximally alien regime. Intermediate values correspond to partial regime shifts. Throughout this work we set $k=1$. One-dimensional loops are the homological dimension most responsive to the cyclical and autocorrelated structure inherent in the embedded trajectory.

\subsection{Computational Aspects}

We compute the persistent homology of the individual point clouds using the Vietoris--Rips filtration, implemented via the \textsc{Ripser} library~\citep{bauer2021ripser}. The mixup barcode is derived by directly implementing Algorithm~1 of~\citet{wagner2024mixup}. We construct the union filtration $\mathrm{VR}(A \cup B)$ a single time, sorting simplices by their filtration values such that \(A\)-simplices precede \(B\)-simplices in the event of a tie. From this, we form two boundary matrices over \(\mathbb{F}_2\): \(B_L\), representing the boundary matrix of \(\mathrm{VR}(A)\) embedded within the index space of the union, and \(B_K\), representing the boundary matrix of the union itself. By permuting the rows to assign lower indices to \(A\)-simplices and subsequently reducing both matrices, the pivot of each killing simplex in the reduced \(B_K\) is forced to point precisely to the \(A\)-simplex whose homology class it annihilates. This procedure yields the canonical matching given by image persistent homology, entirely bypassing the need for either a tolerance parameter or a birth-proximity heuristic. The Vietoris--Rips filtration is truncated at a maximum edge length of $2.5$, and any features that persist beyond this radius are assigned a death time of $2.5$. Because the cost of the union reduction grows rapidly with the number of points, each delay-embedded window (containing $w-(d-1)\tau$ points, i.e., $81$ for SPY and $54$ for Bitcoin at the selected parameters) is subsampled to $28$ points by uniform random sampling without replacement with a fixed seed, the sampled points being kept in temporal order; the reference windows are subsampled in the same way. For each asset, the embedding parameters $(w, d, \tau)$ are selected by maximizing the in-sample Sharpe ratio across a $72$-configuration grid spanning a current window of $w\in\{30,45,60,90\}$ days, an embedding dimension of $d\in\{2,3,4\}$, and a delay of $\tau\in\{1,\dots,6\}$. Throughout this search, the score weights $(w_1,w_2,w_3)$ and thresholds $(\theta_{\mathrm{lo}},\theta_{\mathrm{hi}})$ are held fixed at the values specified in Sections~\ref{sec:stress-def}--\ref{sec:strategy}, with the complete parameter grid detailed in Appendix~\ref{sec:appendix}.

\subsection{Reference and Current Clouds}

The current point cloud $P_t$ is constructed as a daily updated Takens embedding of a trailing $w$-day window of z-scored log prices. In contrast, the reference cloud $P_\mathrm{ref}$ is built a single time using the initial $T+w$ observations, where $T=120$ days is the length of the reference window for both assets, so that the reference slice spans $210$ observations for SPY and $180$ for Bitcoin. We generate a point cloud for every possible window position within this initial segment and retain the central one as the canonical baseline, an approach that avoids conditioning the analysis on a single, potentially atypical market snapshot. The reference cloud $P_\mathrm{ref}$ remains fixed for the remainder of the evaluation. Because the stress score is undefined during the first $T+w$ observations, the strategy is evaluated over this initial period with $S(t)$ artificially set to zero, causing the allocation to coincide exactly with the unmodulated Golden Cross during this warm-up phase.

\subsection{Composite Stress Score}
\label{sec:stress-def}
The stress score is constructed in two stages,
\begin{equation}
\label{eq:stress}
\begin{aligned}
S_{\mathrm{raw}}(t) &= w_1\,\widetilde{W}_1\!\left(D_0^{\mathrm{ref}},\,D_0^t\right)
     +w_2\,\widetilde{MP}_1\!\left(P_\mathrm{ref},\,P_t\right)
     +w_3\,\widetilde{\Delta H}(t),\\[3pt]
S(t) &= \operatorname{clip}\!\left(\frac{1}{30}\sum_{k=0}^{29} S_{\mathrm{raw}}(t-k),\;0,\;1\right),
\end{aligned}
\end{equation}
where $\widetilde{W}_1$ denotes the min--max normalized $1$-Wasserstein distance between the $H_0$ persistence diagrams of $P_\mathrm{ref}$ and $P_t$; $\widetilde{MP}_1$ represents the rescaled $H_1$ mixup percentage introduced in Section~\ref{sec:mixup}; and $\widetilde{\Delta H}(t)=\bigl|H(D_1^t)-H(D_1^{\mathrm{ref}})\bigr|$, normalized in the same manner, captures the divergence in $H_1$ persistence entropy. To ensure the metric remains strictly causal, all three constituent terms---including $MP_1$, despite its inherent $[0,1]$ bounds---are mapped to the $[0,1]$ interval using an expanding min--max normalization over all historical observations up to time $t$. Consequently, the $MP_1$ component utilized in Eq.~\eqref{eq:stress} is this normalized value rather than the raw mixup percentage. The resulting unclipped weighted sum, $S_{\mathrm{raw}}(t)$, is subsequently smoothed via a 30-day trailing moving average and bounded within $[0,1]$ to yield the final score $S(t)$. The weighting vector $(w_1, w_2, w_3) = (0.45, 0.50, 0.05)$ was established a priori to emphasize the primary role of the relational $MP_1$ term, while assigning a secondary, regularizing function to the entropy divergence. The impact of this specific weight selection on the overall results is analyzed in Section~\ref{sec:sensitivity}.

\subsection{Interpretation of the Stress Score}

These three components provide complementary insights into market dynamics. The $H_0$ Wasserstein distance, $\widetilde{W}_1$, detects alterations in the connected component structure, such as the rate at which points aggregate at fine scales, thereby signaling shifts in the local clustering of the embedded trajectory. Conversely, the mixup percentage, $MP_1$, directly quantifies the extent to which the current market state dismantles the reference loop topology. Since it is normalized by feature lifetime, it exhibits strong robustness against outliers. It is important to note that the $MP_1$ term incorporated into $S(t)$ is the expanding min--max rescaled version, meaning that raw $MP_1$ magnitudes cannot be directly compared against the allocation thresholds $\theta_{\mathrm{lo}}$ and $\theta_{\mathrm{hi}}$. Finally, the entropy divergence, $\widetilde{\Delta H}$, monitors the \emph{complexity} of the topological structure, increasing when a proliferation of short-lived features occurs. The predominant weighting assigned to $MP_1$ reflects its capacity to deliver the most acute and timely regime-shift signal, while $\widetilde{W}_1$ supplements this with broader distributional context and $\widetilde{\Delta H}$ serves as a stabilizing regularizer against transient noise spikes.

\section{Strategy Design}
\label{sec:strategy}

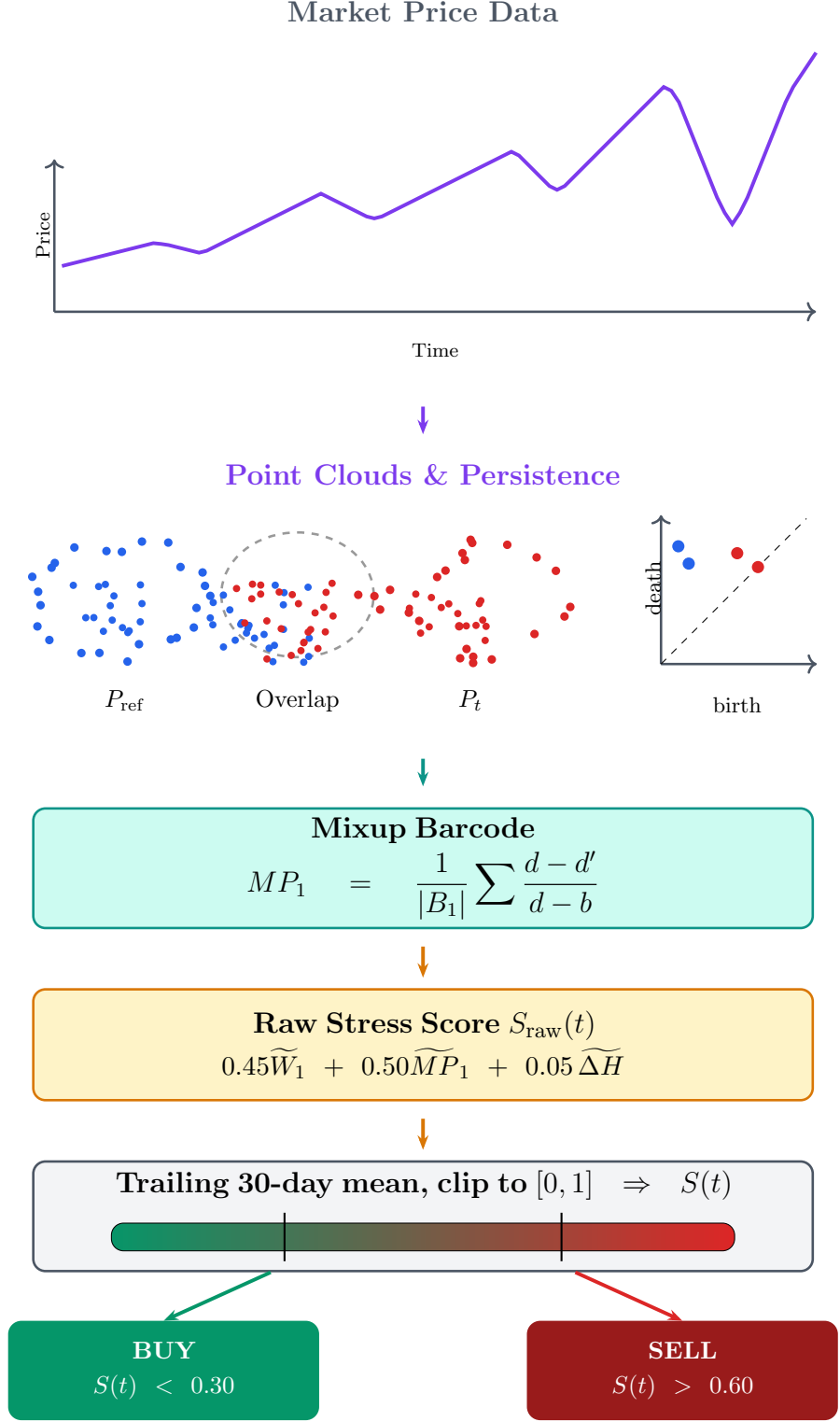
\begin{figure}[H]
\centering
\resizebox{0.72\textwidth}{!}{%
\centering
\begin{tikzpicture}[
node distance=0.55cm and 1.2cm,
every node/.style={font=\fontsize{10}{12}\selectfont},
box/.style={draw, rounded corners=5pt, line width=1pt, align=center, text width=11cm},
smallbox/.style={draw, rounded corners=5pt, line width=1pt, align=center, text width=4.2cm},
arrow/.style={-{Stealth[length=6pt]}, line width=1.5pt},
title/.style={font=\fontsize{13}{15}\selectfont\bfseries}
]

\node[title, color=gaGray] (title1) {Market Price Data};

\node[below=0.15cm of title1] (chart1) {
\begin{tikzpicture}[scale=0.55]
\draw[gaGray, line width=1pt, ->] (0,0) -- (20,0);
\draw[gaGray, line width=1pt, ->] (0,0) -- (0,4);
\node[font=\fontsize{8}{10}\selectfont] at (10,-0.6) {Time};
\node[rotate=90, font=\fontsize{8}{10}\selectfont] at (-0.7,2) {Price};
\draw[gaPurple, line width=1.5pt]
(0.2,1.2) -- (0.4,1.25) -- (0.6,1.3) -- (0.8,1.35) -- (1.0,1.4)
-- (1.2,1.45) -- (1.4,1.5) -- (1.6,1.55) -- (1.8,1.6) -- (2.0,1.65)
-- (2.2,1.7) -- (2.4,1.75) -- (2.6,1.8) -- (2.8,1.78) -- (3.0,1.75)
-- (3.2,1.7) -- (3.4,1.65) -- (3.6,1.6) -- (3.8,1.55) -- (4.0,1.6)
-- (4.2,1.7) -- (4.4,1.8) -- (4.6,1.9) -- (4.8,2.0) -- (5.0,2.1)
-- (5.2,2.2) -- (5.4,2.3) -- (5.6,2.4) -- (5.8,2.5) -- (6.0,2.6)
-- (6.2,2.7) -- (6.4,2.8) -- (6.6,2.9) -- (6.8,3.0) -- (7.0,3.1)
-- (7.2,3.0) -- (7.4,2.9) -- (7.6,2.8) -- (7.8,2.7) -- (8.0,2.6)
-- (8.2,2.5) -- (8.4,2.45) -- (8.6,2.5) -- (8.8,2.6) -- (9.0,2.7)
-- (9.2,2.8) -- (9.4,2.9) -- (9.6,3.0) -- (9.8,3.1) -- (10.0,3.2)
-- (10.2,3.3) -- (10.4,3.4) -- (10.6,3.5) -- (10.8,3.6) -- (11.0,3.7)
-- (11.2,3.8) -- (11.4,3.9) -- (11.6,4.0) -- (11.8,4.1) -- (12.0,4.2)
-- (12.2,4.1) -- (12.4,3.9) -- (12.6,3.7) -- (12.8,3.5) -- (13.0,3.3)
-- (13.2,3.2) -- (13.4,3.3) -- (13.6,3.5) -- (13.8,3.7) -- (14.0,3.9)
-- (14.2,4.1) -- (14.4,4.3) -- (14.6,4.5) -- (14.8,4.7) -- (15.0,4.9)
-- (15.2,5.1) -- (15.4,5.3) -- (15.6,5.5) -- (15.8,5.7) -- (16.0,5.9)
-- (16.2,5.8) -- (16.4,5.5) -- (16.6,5.0) -- (16.8,4.5) -- (17.0,4.0)
-- (17.2,3.5) -- (17.4,3.0) -- (17.6,2.6) -- (17.8,2.3) -- (18.0,2.6)
-- (18.2,3.0) -- (18.4,3.5) -- (18.6,4.0) -- (18.8,4.5) -- (19.0,5.0)
-- (19.2,5.5) -- (19.4,5.9) -- (19.6,6.2) -- (19.8,6.5) -- (20.0,6.8);
\end{tikzpicture}
};

\node[below=0.2cm of chart1] (a1) {};
\draw[arrow, color=gaPurple] (a1.south) -- ($(a1.south)+(0,-0.4)$);

\node[below=0.7cm of a1, title, color=gaPurple] (title2) {Point Clouds \& Persistence};

\node[below=0.15cm of title2] (clouds) {
\begin{tikzpicture}[scale=0.5]
\begin{scope}[shift={(0,0)}]
\foreach \a in {0,15,...,345}{
\pgfmathsetmacro{\x}{2.5*cos(\a) + 0.3*(rand-0.5)}
\pgfmathsetmacro{\y}{1.6*sin(\a) + 0.25*(rand-0.5)}
\fill[gaBlue] (\x,\y) circle (3.5pt);
}
\foreach \i in {1,...,15}{
\pgfmathsetmacro{\x}{1.2*(rand-0.5)}
\pgfmathsetmacro{\y}{0.8*(rand-0.5)}
\fill[gaBlue] (\x,\y) circle (3pt);
}
\node[font=\fontsize{10}{12}\selectfont] at (0,-2.5) {$P_{\text{ref}}$};
\end{scope}
\begin{scope}[shift={(5,0)}]
\draw[black!40, dashed, line width=1pt] (0,0) ellipse (2.2 and 1.8);
\foreach \i in {1,...,25}{
\pgfmathsetmacro{\x}{1.5*(rand-0.5) - 0.3}
\pgfmathsetmacro{\y}{1.3*(rand-0.5)}
\fill[gaBlue] (\x,\y) circle (3pt);
}
\foreach \i in {1,...,25}{
\pgfmathsetmacro{\x}{1.5*(rand-0.5) + 0.3}
\pgfmathsetmacro{\y}{1.3*(rand-0.5)}
\fill[gaCoral] (\x,\y) circle (3pt);
}
\node[font=\fontsize{10}{12}\selectfont] at (0,-2.5) {Overlap};
\end{scope}
\begin{scope}[shift={(10,0)}]
\foreach \a in {0,15,...,345}{
\pgfmathsetmacro{\x}{2.2*cos(\a) + 0.7*cos(3*\a) + 0.3*(rand-0.5)}
\pgfmathsetmacro{\y}{1.5*sin(\a) + 0.5*sin(2*\a) + 0.25*(rand-0.5)}
\fill[gaCoral] (\x,\y) circle (3.5pt);
}
\foreach \i in {1,...,15}{
\pgfmathsetmacro{\x}{1.2*(rand-0.5)}
\pgfmathsetmacro{\y}{0.9*(rand-0.5)}
\fill[gaCoral] (\x,\y) circle (3pt);
}
\node[font=\fontsize{10}{12}\selectfont] at (0,-2.5) {$P_t$};
\end{scope}
\begin{scope}[shift={(15.5,0)}]
\draw[gaGray, line width=1pt, ->] (0,-2) -- (0,2.3);
\draw[gaGray, line width=1pt, ->] (0,-2) -- (4.5,-2);
\node[font=\fontsize{9}{11}\selectfont, rotate=90] at (-0.7,0.2) {death};
\node[font=\fontsize{9}{11}\selectfont] at (2.2,-2.7) {birth};
\draw[dashed] (0,-2) -- (4.2,2.2);
\fill[gaBlue]  (0.5,1.4) circle (5pt);
\fill[gaBlue]  (0.8,0.9) circle (5pt);
\fill[gaCoral] (2.2,1.2) circle (5pt);
\fill[gaCoral] (2.8,0.8) circle (5pt);
\end{scope}
\end{tikzpicture}
};

\node[below=0.15cm of clouds] (a2) {};
\draw[arrow, color=gaTeal] (a2.south) -- ++(0,-0.4);

\node[below=0.7cm of a2, box, fill=gaTealLight, draw=gaTeal, minimum height=1.4cm] (mixup) {
{\fontsize{13}{15}\selectfont\bfseries Mixup Barcode}\\[2pt]
{\fontsize{12}{14}\selectfont $\displaystyle MP_1 = \frac{1}{|B_1|}\sum \frac{d-d'}{d-b}$}
};

\draw[arrow, color=gaAmber] ($(mixup.south)+(0,-0.25)$) -- ++(0,-0.45);

\node[below=0.85cm of mixup, box, fill=gaAmberLight, draw=gaAmber, minimum height=1.6cm] (stress) {
{\fontsize{13}{15}\selectfont\bfseries Raw Stress Score $S_{\mathrm{raw}}(t)$}\\[2pt]
{\fontsize{11}{13}\selectfont $0.45\widetilde{W}_1 + 0.50\widetilde{MP}_1 + 0.05\,\widetilde{\Delta H}$}
};

\draw[arrow, color=gaAmber] ($(stress.south)+(0,-0.25)$) -- ++(0,-0.45);

\node[below=0.85cm of stress, box, fill=gaGrayLight, draw=gaGray, minimum height=1.5cm] (smooth) {
{\fontsize{12}{14}\selectfont\bfseries Trailing 30-day mean, clip to $[0,1]$ $\;\Rightarrow\;$ $S(t)$}\\[4pt]
\begin{tikzpicture}
\shade[left color=gaGreen, right color=gaCoral] (0,0) rectangle (9,0.4);
\draw (0,0) rectangle (9,0.4);
\draw[thick] (2.5,-0.15)--(2.5,0.55);
\draw[thick] (6.5,-0.15)--(6.5,0.55);
\end{tikzpicture}
};

\node[below left=0.7cm and 1.5cm of smooth.south, smallbox,
fill=gaGreen, draw=gaGreen, minimum height=1.4cm] (buy) {
{\bfseries\color{white} BUY}\\[2pt]
{\color{white} $S(t)<0.30$}
};

\node[below right=0.7cm and 1.5cm of smooth.south, smallbox,
fill=gaCoralDark, draw=gaCoralDark, minimum height=1.4cm] (sell) {
{\bfseries\color{white} SELL}\\[2pt]
{\color{white} $S(t)>0.60$}
};

\draw[arrow, color=gaGreen] (smooth.south) ++(-2.2,0) -- (buy.north);
\draw[arrow, color=gaCoral] (smooth.south) ++(2.2,0) -- (sell.north);

\end{tikzpicture}
}
\caption{Overview of the topology-aware trading framework. Raw market price data are transformed into point clouds via the Takens delay embedding, enabling the computation of persistent homology and mixup barcodes. The raw stress score is subsequently smoothed using a 30-day trailing moving average to yield $S(t)$, which ultimately dictates the position sizing.}
\label{fig:pipeline}
\end{figure}

We integrate the mixup barcode (Section~\ref{sec:mixup}) and the stress score (Section~\ref{sec:stress-def}) into the comprehensive trading pipeline illustrated in Figure~\ref{fig:pipeline}, an architecture that enforces a clear separation of concerns. A conventional trend filter dictates the \emph{direction} of market exposure, while the topological stress score determines its \emph{magnitude}.  In essence, the trend component governs the decision to participate in the market, whereas the topological component calibrates the size of the position.

The pipeline proceeds sequentially from top to bottom, beginning with the extraction of a trailing window of z-scored log prices for each trading day (\emph{Stage~1}), a process that requires no distributional assumptions. In \emph{Stage~2}, each window is mapped into $\mathbb{R}^d$ via the Takens embedding, where the reference cloud $P_\mathrm{ref}$ typically forms a smooth, low-dimensional arc characteristic of a tranquil market regime, whereas the current cloud $P_t$ exhibits a more distorted geometry. The corresponding persistence diagrams on the right illustrate how the $H_1$ topological features of $P_\mathrm{ref}$ (depicted in blue) are drawn toward the diagonal and effectively truncated when the current state $P_t$ (shown in coral) is introduced. \emph{Stage~3} condenses this topological disruption into the scalar metric $MP_1=\tfrac{1}{|B_1|}\sum_{i\in B_1}(d_i-d_i')/(d_i-b_i)$, which represents the mean fractional lifetime lost by the reference loops; the defining characteristic of this metric is its \emph{relational} nature, which quantifies the disruption of one state by another rather than evaluating them in isolation, thereby setting it apart from conventional TDA indicators \citep{wagner2024mixup,yadav2025missing}. Finally, \emph{Stage~4} integrates $MP_1$ with $\widetilde{W}_1$ and $\widetilde{\Delta H}$ to compute $S_{\mathrm{raw}}(t)$, which is then smoothed and bounded to produce the final stress score $S(t)$ according to Eq.~\eqref{eq:stress}. The BUY and SELL indicators at the bottom represent the saturated extremes of the continuous, trend-gated position sizing mechanism detailed subsequently.

\subsection{Golden Cross Baseline}

The \emph{Golden Cross} strategy initiates a long position when the 50-day simple moving average (SMA) surpasses the 200-day SMA, and exits the market upon the inverse \emph{Death Cross} signal. Because it is among the most widely used trend-following rules, we adopt this approach as our primary benchmark. In its conventional formulation, the strategy maintains a binary exposure profile, remaining either fully invested in the underlying asset or entirely in cash.

\subsection{Topology-Aware Position Sizing}

We set the target portfolio weight
\[
\alpha(t)=
\begin{cases}
1.0 & \text{if } S(t)<\theta_{\mathrm{lo}}=0.30,\\[3pt]
\dfrac{\theta_{\mathrm{hi}}-S(t)}{\theta_{\mathrm{hi}}-\theta_{\mathrm{lo}}}
   & \text{if } \theta_{\mathrm{lo}}\leq S(t)\leq\theta_{\mathrm{hi}},\\[5pt]
0.0 & \text{if } S(t)>\theta_{\mathrm{hi}}=0.60.
\end{cases}
\]
A long position is initiated exclusively on the day a Golden Cross signal occurs, provided that \(S(t) < \theta_{\mathrm{hi}}\) on that specific day. Any Golden Cross triggered while \(S(t) \geq \theta_{\mathrm{hi}}\) is ignored, leaving the strategy in a flat position until the subsequent crossover event. During active investment periods, the position size is dynamically adjusted on a daily basis according to the allocation function \(\alpha(t)\) defined previously. Once $\alpha(t)$ reaches zero because $S(t)\geq\theta_{\mathrm{hi}}$, the position is
not reopened if $S(t)$ later falls back; the strategy stays flat until the next Golden
Cross. A Death Cross forces $\alpha(t)=0$ unconditionally. The threshold parameters \((\theta_{\mathrm{lo}},\theta_{\mathrm{hi}})=(0.30,0.60)\) were established a priori to facilitate a smooth, continuous de-risking process. Specifically, the strategy maintains full exposure when the current market geometry closely mirrors the reference state (indicating a tranquil, trend-consistent regime), applies linear de-risking as $S(t)$ increases and structural risk accumulates, and executes a complete exit once $S(t)>\theta_{\mathrm{hi}}$, a regime where traditional trend signals are deemed unreliable. We refer to the resulting strategy as \emph{TDA Stress GC}.

\subsection{Transaction Costs}

The backtesting framework incorporates a one-way transaction cost of 5 basis points applied to the absolute daily change in portfolio allocation, reflecting a daily rebalancing schedule. To realistically account for execution delays, both the position adjustments and their corresponding transaction costs are implemented with a one-day lag, ensuring that the trading fee is deducted on the precise day the new exposure is first actively held.

\section{Empirical Results}
\label{sec:results}

\subsection{Data and Setup}
Our empirical evaluation focuses on two distinct asset classes. The S\&P~500, represented by the SPY ETF (daily adjusted closing prices from January~2010 to December~2024, comprising 3,771 trading days), and Bitcoin, represented by the BTC/USD pair (from January~2017 to December~2024, comprising 2,919 days). For each asset, the optimal embedding parameters were identified by maximizing the Sharpe ratio across the 72-configuration grid detailed in Section~\ref{sec:mixup}, yielding $(w=90, d=4, \tau=3)$ for SPY and $(w=60, d=4, \tau=2)$ for Bitcoin. Throughout this analysis, the score weights, decision thresholds, and window lengths remained strictly fixed across both markets to ensure a consistent evaluation framework. Since the embedding parameters are optimized on the identical dataset used to report the final performance metrics, the headline results represent in-sample figures, with the complete parameter grid documented in Appendix~\ref{sec:appendix}.

\subsection{Metrics}

To comprehensively evaluate performance, we report a suite of standard financial metrics: the compound annual growth rate (CAGR); the Sharpe ratio, computed as the annualized excess return scaled by volatility with a risk-free rate of $0$; the Sortino ratio, which penalizes only downside deviation; the maximum drawdown (Max DD); the Calmar ratio, defined as the ratio of CAGR to $|\text{Max DD}|$; and the total frequency of trading signals, encompassing both position entries and exits. All annualized quantities are computed with $252$ periods per year for both assets. Because Bitcoin trades every calendar day ($365$ observations per year), its annualized returns and ratios are understated relative to a $365$-day convention: the Sharpe and Sortino ratios by a uniform factor of $\sqrt{365/252}\approx 1.20$, and the annualized return and Calmar ratio nonlinearly (for example, Buy \& Hold on Bitcoin corresponds to a calendar-year CAGR of about $76\%$ rather than the $47.6\%$ reported in Table~\ref{tab:btc}). Since the convention is common to all strategies, the comparisons and rankings below are unaffected, and the maximum drawdown does not depend on it.

\subsection{S\&P 500 Results}
\label{sec:sp500-results}
\begin{table}[t]
\centering
\caption{Full-sample performance on the S\&P\,500 (Jan~2010--Dec~2024), net of transaction costs. Buy \& Hold enters once and never exits.}
\label{tab:sp500}
\begin{tabular}{lcccccc}
\toprule
Strategy & Return & Sharpe & Sortino & Max DD & Calmar & Signals \\
\midrule
TDA Stress GC & 0.096 & 1.116 & 1.198 & $-$0.164 & 0.584 & 13 \\
Buy \& Hold   & 0.137 & 0.838 & 1.026 & $-$0.337 & 0.406 & --  \\
Golden Cross  & 0.095 & 0.733 & 0.760 & $-$0.337 & 0.282 & 13 \\
\bottomrule
\end{tabular}
\end{table}

The TDA Stress GC returns $9.6\%$ annually, essentially matching the standard Golden Cross
($9.5\%$) while raising the Sharpe ratio by $\sim$52\% ($1.116$ vs.\ $0.733$) and halving
the maximum drawdown ($-16.4\%$ vs.\ $-33.7\%$). The Calmar ratio rises to $0.584$ from
$0.282$. Buy \& Hold earns more raw return ($13.7\%$) but with more than double the
drawdown and a lower Sharpe ($0.838$). The strategy thus dominates both benchmarks on
every risk-adjusted metric.

\subsection{Bitcoin Results}
\label{sec:btc-results}
\begin{table}[t]
\centering
\caption{Full-sample performance on Bitcoin (Jan~2017--Dec~2024), net of transaction costs. Buy \& Hold enters once and never exits.}
\label{tab:btc}
\begin{tabular}{lcccccc}
\toprule
Strategy & Return & Sharpe & Sortino & Max DD & Calmar & Signals \\
\midrule
TDA Stress GC & 0.196 & 1.238 & 1.223 & $-$0.184 & 1.065 & 16 \\
Buy \& Hold   & 0.476 & 0.954 & 1.298 & $-$0.834 & 0.570 & --  \\
Golden Cross  & 0.349 & 0.871 & 0.886 & $-$0.692 & 0.505 & 15 \\
\bottomrule
\end{tabular}
\end{table}

Improvements are larger on Bitcoin: Sharpe rises to $1.238$ vs.\ $0.871$ for the Golden Cross (a $42\%$ gain), with Max DD cut from $-69.2\%$ to $-18.4\%$ and Calmar reaching $1.065$, far above the Golden Cross ($0.505$) and Buy \& Hold ($0.570$). Buy \& Hold's higher raw return ($47.6\%$ vs.\ $19.6\%$) comes at the cost of an $-83.4\%$ drawdown. Its slightly higher Sortino ratio ($1.298$ vs.\ $1.223$) reflects Bitcoin's extreme upside volatility rather than downside protection. The sharper gains are consistent with the pronounced crash dynamics of cryptocurrency markets \citep{gidea2020topological,mukhia2024complex}.

\subsection{Stress Score Dynamics and Sensitivity}
\label{sec:sensitivity}

\begin{figure}[hbt]
    \centering
    \includegraphics[width=13cm, keepaspectratio]{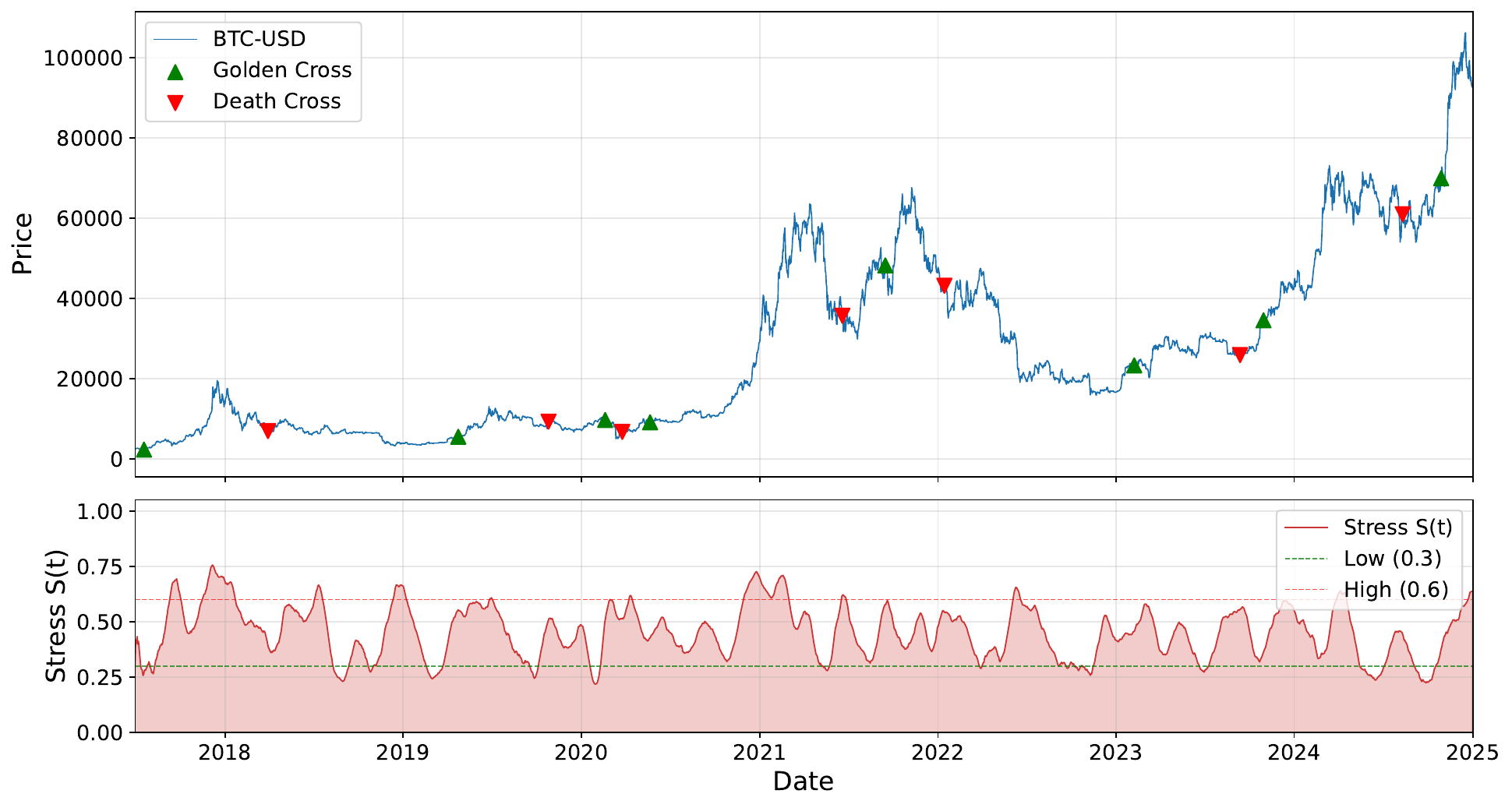}
    \caption{Bitcoin price trajectory (top panel) annotated with Golden and Death Cross signals, alongside the corresponding topological stress score $S(t)$ (bottom panel). The dashed horizontal lines mark the position-sizing thresholds: full exposure is maintained when $S(t) < \theta_{\mathrm{lo}} = 0.30$, the allocation scales linearly for intermediate values between $0.30$ and $0.60$, and the strategy completely liquidates its position once $S(t)$ exceeds $\theta_{\mathrm{hi}} = 0.60$.}
    \label{fig:stress}
\end{figure}

Figure~\ref{fig:stress} plots $S(t)$ against the Bitcoin price. The score varies
smoothly --- a consequence of the 30-day trailing mean applied to $S_{\mathrm{raw}}$ ---
and spends most of the sample ($78\%$ of days) between the two thresholds, exceeding
$0.60$ on $11\%$ of days and falling below $0.30$ on $12\%$. It therefore acts mainly as
a continuous de-risking dial rather than a binary regime switch: average exposure over
the sample is $14\%$ ($26\%$ while the Golden Cross trend filter is active). Elevated
readings are not confined to bear markets. For example, $S(t)$ exceeded $0.60$ on $44\%$
of days between October 2020 and April 2021, a period of strong price appreciation.

Extensive robustness sweeps around the baseline configuration (yielding a Sharpe ratio of $1.116$ for SPY and $1.238$ for Bitcoin)    show that the baseline is the best configuration in both window sweeps, and that
performance degrades unevenly away from it. Varying the reference window $T\in\{60,\ldots,180\}$ days yields a Sharpe range of $[0.13,\,1.12]$ for SPY and $[0.60,\,1.24]$ for Bitcoin, while sweeping the current window $w\in\{20,\ldots,90\}$ produces $[0.31,\,1.12]$ and $[0.65,\,1.24]$, respectively; for SPY, performance collapses at long reference windows ($T\geq150$: Sharpe $0.28$ and $0.13$), and the baseline $w=90$ is the largest current window tested; for Bitcoin, the baseline is a sharp peak in both sweeps, with its worst reference window at $T=60$ (Sharpe $0.60$). Perturbing the component weights $(w_1,w_2,w_3)$ by $\pm10\%$ alters the Sharpe ratio by a maximum of $0.017$ for SPY but up to $0.224$ for Bitcoin, demonstrating that the precise composition of the stress score is more critical for the highly volatile asset. Finally, across the threshold grid $(\theta_{\mathrm{lo}},\theta_{\mathrm{hi}})\in\{0.20,\ldots,0.40\}\times\{0.50,\ldots,0.75\}$, the proposed strategy outperforms the standard Golden Cross across every parameter combination for SPY (Sharpe $[0.77,\,1.12]$) and in $83\%$ of the combinations for Bitcoin (Sharpe $[0.66,\,1.25]$).

\section{Discussion}
\label{sec:discussion}

\subsection{Why the Mixup Barcode Works}

The regime information captured by $MP_1$ is fundamentally geometric in nature. During calm market conditions, the embedded log-price window forms a smooth, low-dimensional arc in $\mathbb{R}^4$ characterized by stable $H_1$ topological features. However, as market stress accumulates, the underlying attractor begins to fray and short-lived features proliferate, a degradation that $MP_1$ quantifies directly. Because this geometric disruption is not captured by conventional moving-average crossovers, the topological stress score and the trend filter provide complementary information, as argued in Section~\ref{sec:introduction}.

\subsection{Limitations}

The union reduction underlying Algorithm~1 exhibits rapid computational scaling with respect to the number of points, necessitating the subsampling of each window to 28 points. This discretization coarsens $MP_1$, which is subsequently averaged over approximately three non-trivial loops per window. A natural alternative that avoids this bottleneck is the mixup Euler characteristic profile, which measures the interaction of two point clouds through the Euler characteristic of the intersection of their ball unions; it is computed by a single sorted Alpha-complex sweep with no persistence reduction \citep{kawamura2026intersection} and has already been used to detect regime transitions in dynamical systems \citep{majhi2026regime}, so it could be applied to full, unsubsampled windows. Furthermore, the proposed framework relies on several hyperparameters, viz., embedding dimension, lag, window lengths, weights, and thresholds. Our sensitivity analysis shows that performance depends materially on these choices, and the development of principled, data-adaptive parameter selection remains an open problem. Additionally, the reference point cloud is held fixed throughout the entire evaluation period, a design choice that may prove overly rigid during periods of rapid structural market transformation. We also acknowledge that the embedding parameters are optimized by maximizing the Sharpe ratio on the identical sample used to report the final performance metrics. Nevertheless, Appendix~\ref{sec:appendix} shows that the Sharpe ratio is positive across nearly the entire grid (71 of 72 Bitcoin configurations and 70 of 72 SPY configurations) and that the drawdown reduction is robust (shallower than the Golden Cross in 71 of 72 and 64 of 72 configurations, respectively). The Sharpe advantage over the Golden Cross is not: only 22 of 72 Bitcoin and 26 of 72 SPY configurations exceed it. The headline Sharpe ratios should therefore be read as an upper bound, and a walk-forward out-of-sample validation is left to future work. Finally, while we do not explicitly model institutional-scale market impact, the exceptionally low trading frequency (only 7 entries over 15 years for SPY and 8 over 8 years for Bitcoin) suggests that such frictions would be negligible at typical fund sizes.

\section{Conclusion}
\label{sec:conclusion}

In this work, we have introduced a topology-aware framework for financial decision-making anchored by \emph{mixup barcodes}. The foundation of this approach is the mixup percentage, $MP_1$, which precisely quantifies the extent to which the topological structure of a reference market state is dismantled by the current state. By integrating this metric with the Wasserstein distance and persistence-entropy divergence into a unified composite stress score, we effectively transform the traditional binary Golden Cross signal into a continuous, risk-sensitive position sizing mechanism. Evaluated on the S\&P\,500 and Bitcoin, the proposed strategy consistently enhances the Sharpe ratio, maximum drawdown, and Calmar ratio relative to both the Buy \& Hold baseline and the standard Golden Cross, while also improving the Sortino ratio over the Golden Cross. The primary conceptual contribution of this research is the introduction of a \emph{relational} topological measure that captures the mutual disruption between market states rather than evaluating them in isolation, a paradigm we anticipate will prove highly valuable for multi-asset regime detection, dynamic portfolio construction, and real-time risk monitoring.

\paragraph{Future Directions.}
Promising avenues for future research include the development of multi-asset portfolios designed to track topological interactions across different market sectors; during financial crises, typically independent assets tend to comove, an event that would collapse a jointly embedded point cloud into a lower-dimensional subspace and obliterate the loops of a reference cloud, representing precisely the type of structural disruption the mixup barcode is engineered to detect. Additional directions involve exploring higher-dimensional homology, such as $H_2$ voids, within large-panel embeddings. Furthermore, future work should encompass a comprehensive walk-forward out-of-sample validation accompanied by formal significance testing (exact tests for topological interaction of the kind developed in \citet{majhi2026certified}, or the permutation test with a built-in null that accompanies the mixup Euler characteristic profile \citep{majhi2026regime}, are natural starting points), a direct comparison of $MP_1$ with the mixup Euler characteristic profile as the relational component of the stress score, the implementation of streaming mixup-barcode computations for intraday trading applications, the integration of mixup-derived features into gradient-boosting or deep learning architectures, and the derivation of theoretical stability guarantees for $MP_1$, in the spirit of those established for the intersection Euler characteristic profile \citep{kawamura2026intersection}, to formally characterize the conditions under which it serves as a consistent statistical test for regime shifts.

\section*{Acknowledgments}
We gratefully acknowledge computational resources provided by Montana Technological University. We also acknowledge NIT Sikkim for awarding a doctoral fellowship to Buddha Nath Sharma.

\bibliographystyle{plainnat}
\bibliography{references}

\appendix
\section{Embedding-Parameter Grid Search}
\label{sec:appendix}
\begin{figure}[p]
\centering
\begin{subfigure}{\textwidth}
  \centering
  \includegraphics[width=\textwidth]{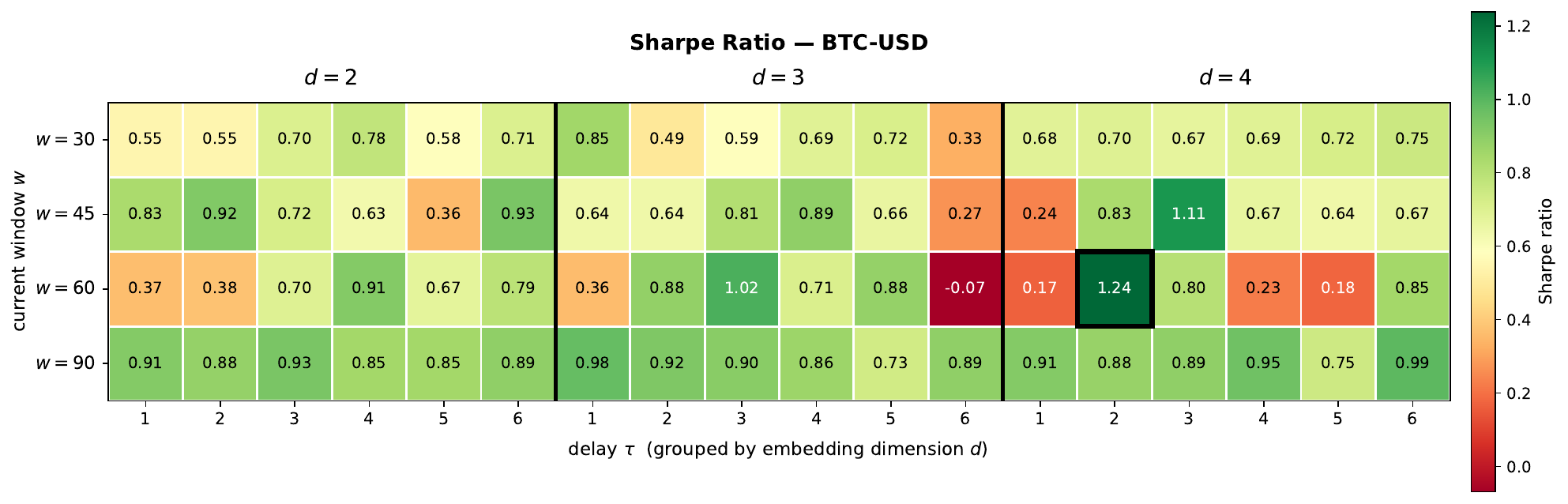}
  \caption{Bitcoin (BTC/USD). Sharpe-maximizing cell: $w=60$, $d=4$, $\tau=2$.}
\end{subfigure}

\vspace{1.2em}

\begin{subfigure}{\textwidth}
  \centering
  \includegraphics[width=\textwidth]{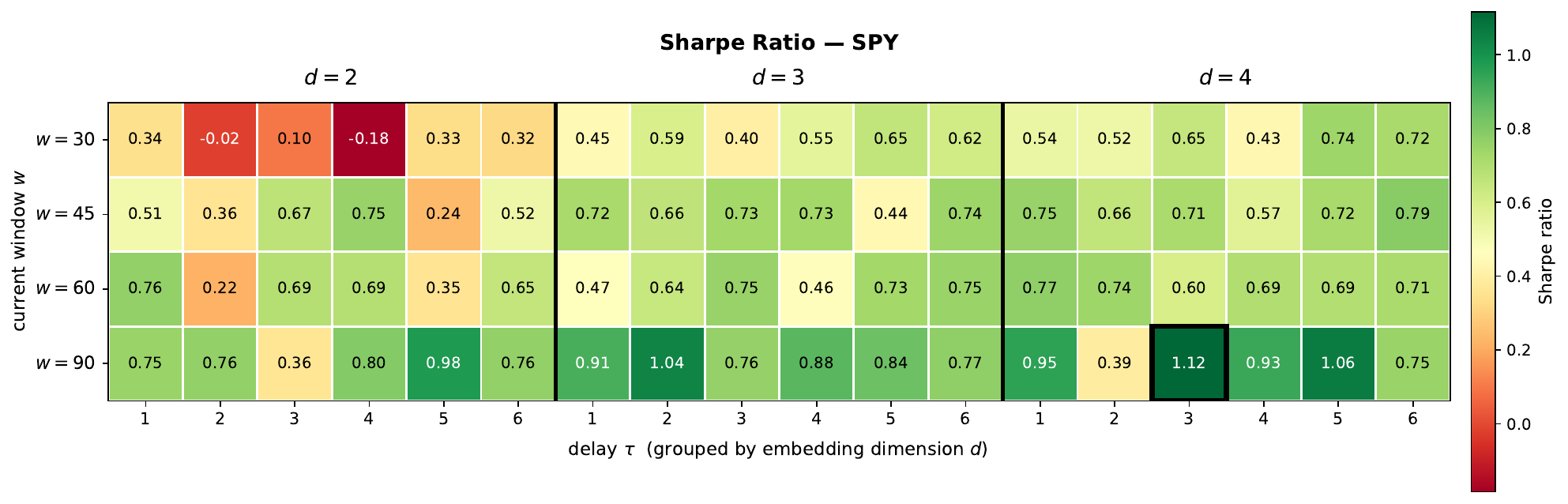}
  \caption{S\&P 500 (SPY). Sharpe-maximizing cell: $w=90$, $d=4$, $\tau=3$.}
\end{subfigure}
\caption{This figure illustrates the Sharpe ratio achieved across all configurations of the Takens-embedding grid search ($72$ distinct parameter combinations per asset), with all returns calculated net of transaction costs. The columns are organized by embedding dimension, while the time delay varies sequentially within each respective group. For each asset, the specific configuration that maximizes the Sharpe ratio and is utilized in the primary analysis is highlighted with a black outline.}
\label{fig:grid-heatmap}
\end{figure}

To confirm that the documented performance does not merely reflect an artifact of a specific embedding selection, we assessed the strategy across the exhaustive grid of Takens parameters: a current window of $w\in\{30,45,60,90\}$ days, an embedding dimension of $d\in\{2,3,4\}$, and a delay of $\tau\in\{1,\dots,6\}$. This yields $72$ distinct configurations per asset, while the score weights $(w_1,w_2,w_3)$ and decision thresholds $(\theta_{\mathrm{lo}},\theta_{\mathrm{hi}})$ were maintained at the fixed values established in Sections~\ref{sec:stress-def}--\ref{sec:strategy}. Figure~\ref{fig:grid-heatmap} visualizes the resulting Sharpe ratios for every configuration across both assets. For Bitcoin, the Sharpe ratio is positive on $71$ of the $72$ configurations (the
sole exception, $w=60$, $d=3$, $\tau=6$, has Sharpe $-0.07$), but only $22$ exceed the
Golden Cross Sharpe of $0.871$ and $6$ exceed Buy \& Hold ($0.954$). The surface is
rugged rather than smooth: for $w=60$, $d=4$, the Sharpe ratio at
$\tau=1,\dots,6$ is $0.17$, $1.24$, $0.80$, $0.23$, $0.18$, $0.85$. The more robust
benefit is drawdown: $71$ of $72$ configurations have a shallower maximum drawdown than
the Golden Cross ($-69.2\%$) and all $72$ than Buy \& Hold ($-83.4\%$), with a median of
$-44.6\%$. For the S\&P\,500, the Sharpe ratio is positive on $70$ of $72$
configurations (the exceptions, $w=30$, $d=2$, $\tau=2$ and $w=30$, $d=2$, $\tau=4$,
have Sharpe $-0.02$ and $-0.18$ respectively); $26$ exceed the Golden Cross ($0.733$)
and $8$ exceed Buy \& Hold ($0.838$), while $64$ configurations have a shallower maximum drawdown than the Golden Cross and Buy \& Hold ($-33.7\%$ for both) and the remaining $8$ match it, with a median of $-20.6\%$. The specific configuration adopted in the primary analysis corresponds to the Sharpe-maximizing cell within each asset's respective grid and is highlighted with a black outline in Figure~\ref{fig:grid-heatmap}. For Bitcoin, the grid-average Sharpe ratio stands at $0.711$ (with a median of $0.727$) compared to the selected peak of $1.238$; for the S\&P\,500, the grid-average is $0.627$ (median $0.690$) relative to the chosen maximum of $1.116$ achieved at $(w=90, d=4, \tau=3)$. In both instances, the observed discrepancy between the grid average and the optimized selection is attributable to the inherent selection effect.

\FloatBarrier

\begin{figure}[p]
\centering
\begin{subfigure}{\textwidth}
  \centering
  \includegraphics[width=\textwidth]{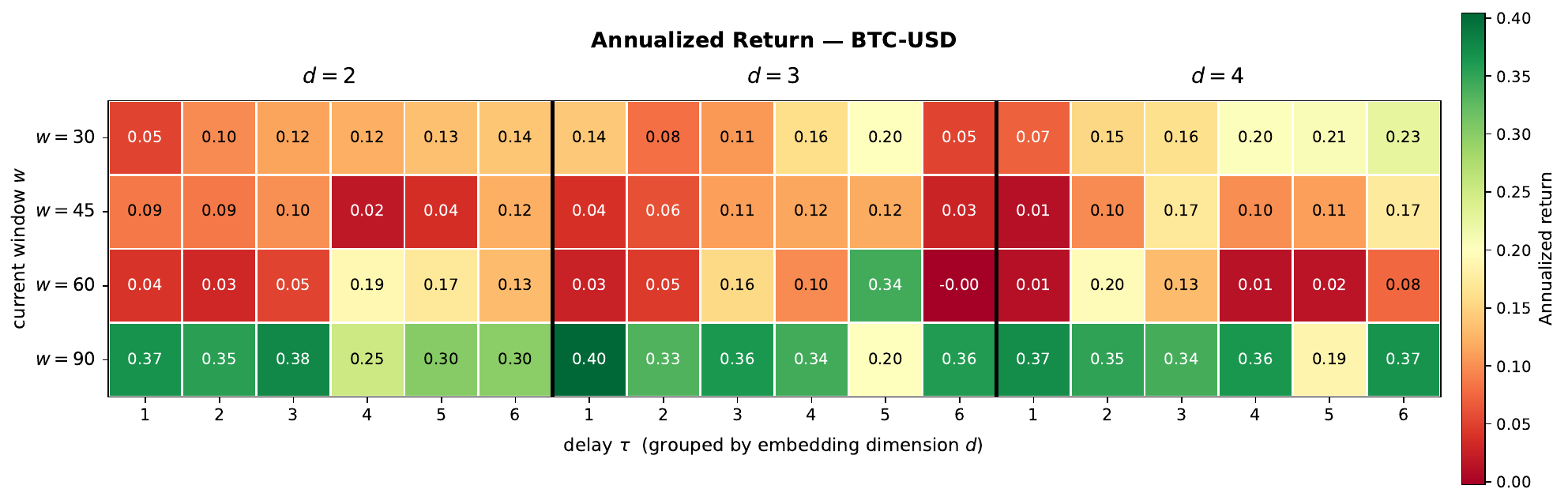}
  \caption{Bitcoin (BTC/USD).}
\end{subfigure}

\vspace{1.2em}

\begin{subfigure}{\textwidth}
  \centering
  \includegraphics[width=\textwidth]{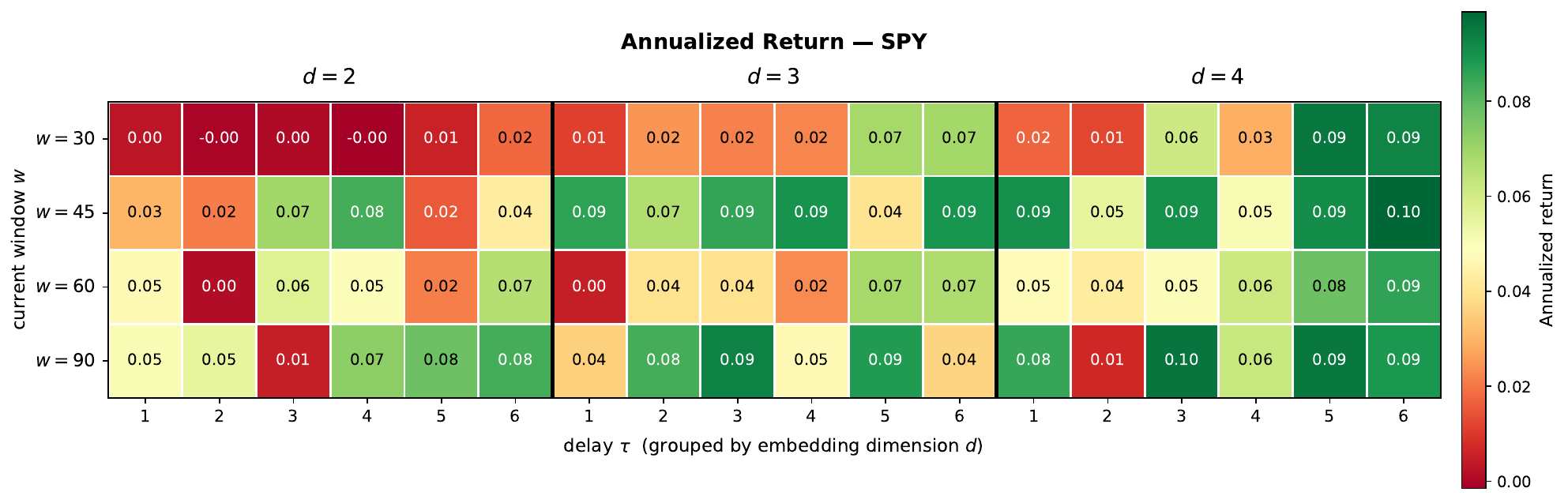}
  \caption{S\&P 500 (SPY).}
\end{subfigure}
\caption{Annualized return for every configuration of the Takens-embedding
grid search (color scaled per asset); layout as in
Figure~\ref{fig:grid-heatmap}.}
\label{fig:grid-ann-return}
\end{figure}
\FloatBarrier

\begin{figure}[p]
\centering
\begin{subfigure}{\textwidth}
  \centering
  \includegraphics[width=\textwidth]{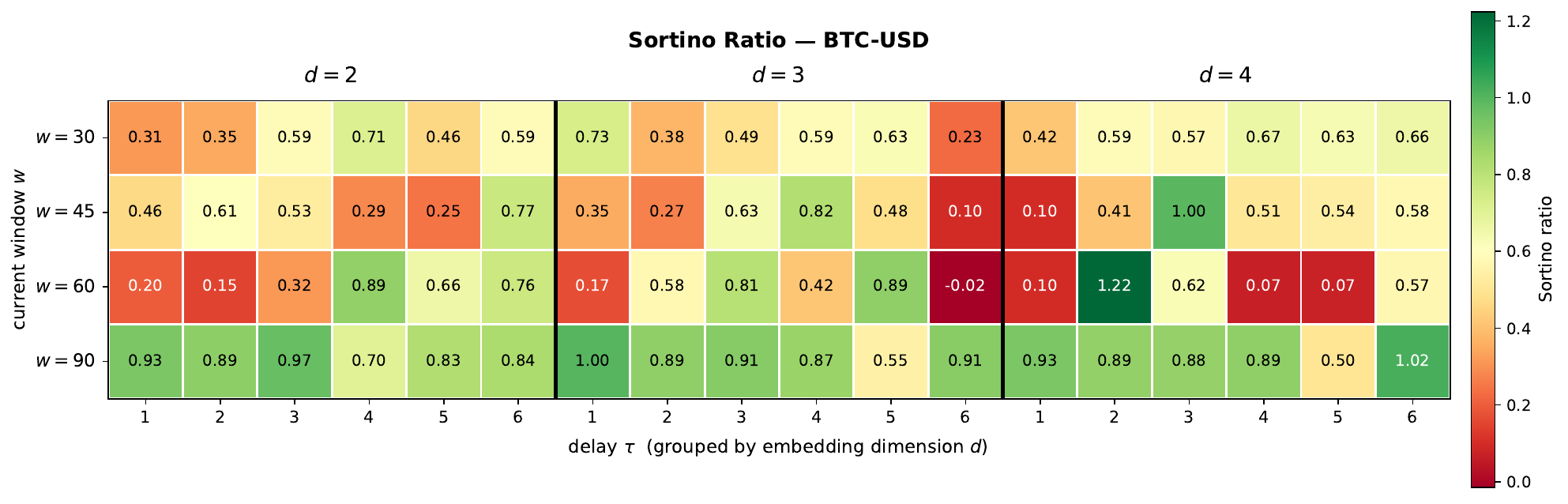}
  \caption{Bitcoin (BTC/USD).}
\end{subfigure}

\vspace{1.2em}

\begin{subfigure}{\textwidth}
  \centering
  \includegraphics[width=\textwidth]{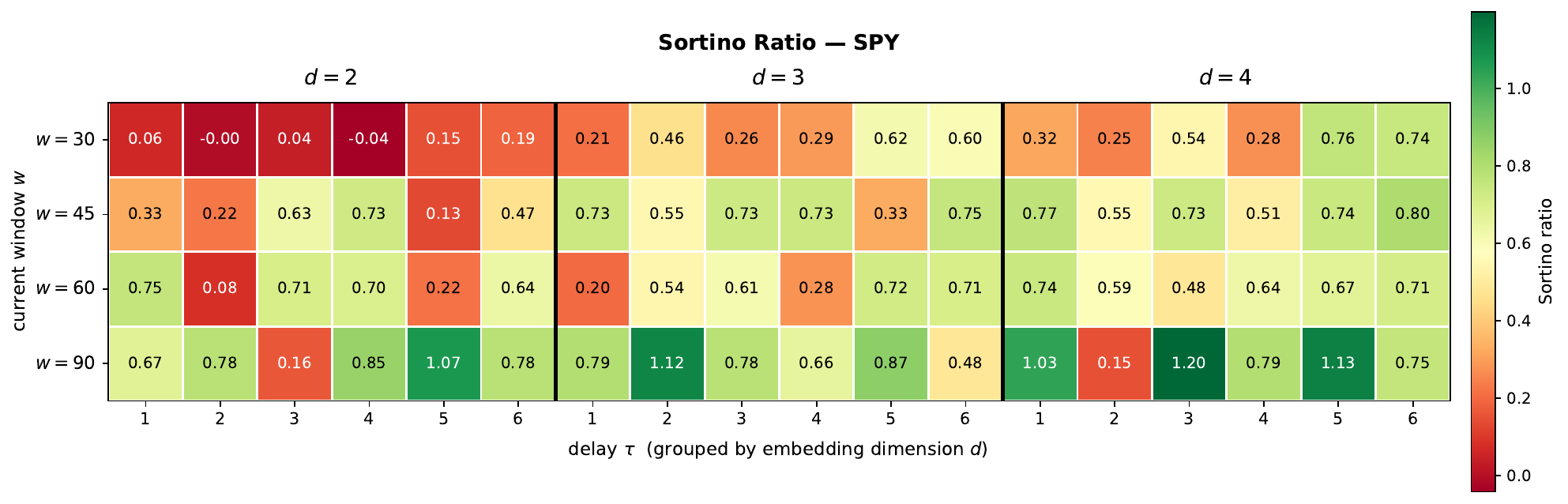}
  \caption{S\&P 500 (SPY).}
\end{subfigure}
\caption{Sortino ratio for every configuration of the Takens-embedding grid
search; layout as in Figure~\ref{fig:grid-heatmap}.}
\label{fig:grid-sortino}
\end{figure}
\FloatBarrier

\begin{figure}[p]
\centering
\begin{subfigure}{\textwidth}
  \centering
  \includegraphics[width=\textwidth]{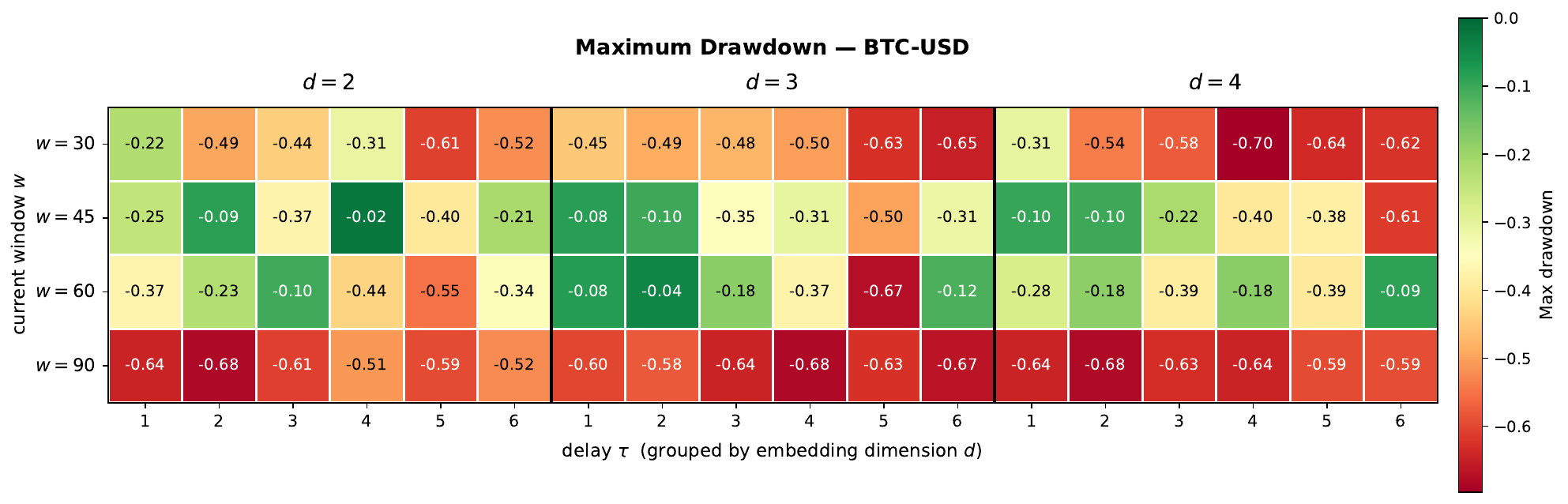}
  \caption{Bitcoin (BTC/USD).}
\end{subfigure}

\vspace{1.2em}

\begin{subfigure}{\textwidth}
  \centering
  \includegraphics[width=\textwidth]{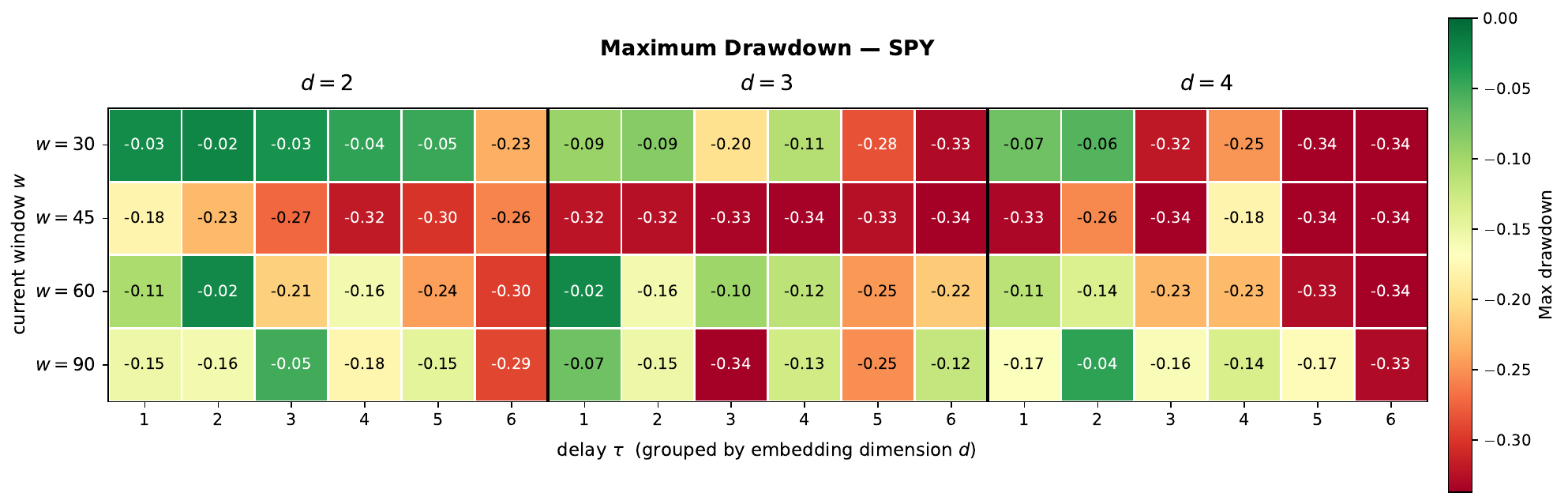}
  \caption{S\&P 500 (SPY).}
\end{subfigure}
\caption{Maximum drawdown for every configuration of the Takens-embedding
grid search (color scaled per asset; values nearer zero are better); layout
as in Figure~\ref{fig:grid-heatmap}.}
\label{fig:grid-max-dd}
\end{figure}
\FloatBarrier

\begin{figure}[p]
\centering
\begin{subfigure}{\textwidth}
  \centering
  \includegraphics[width=\textwidth]{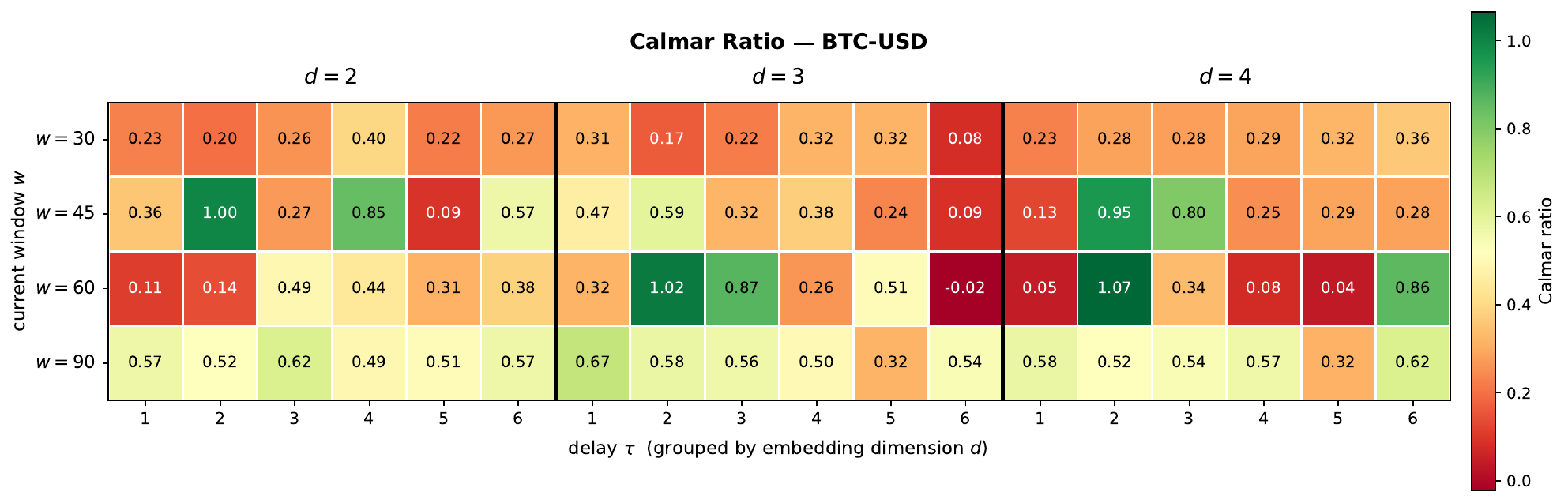}
  \caption{Bitcoin (BTC/USD).}
\end{subfigure}

\vspace{1.2em}

\begin{subfigure}{\textwidth}
  \centering
  \includegraphics[width=\textwidth]{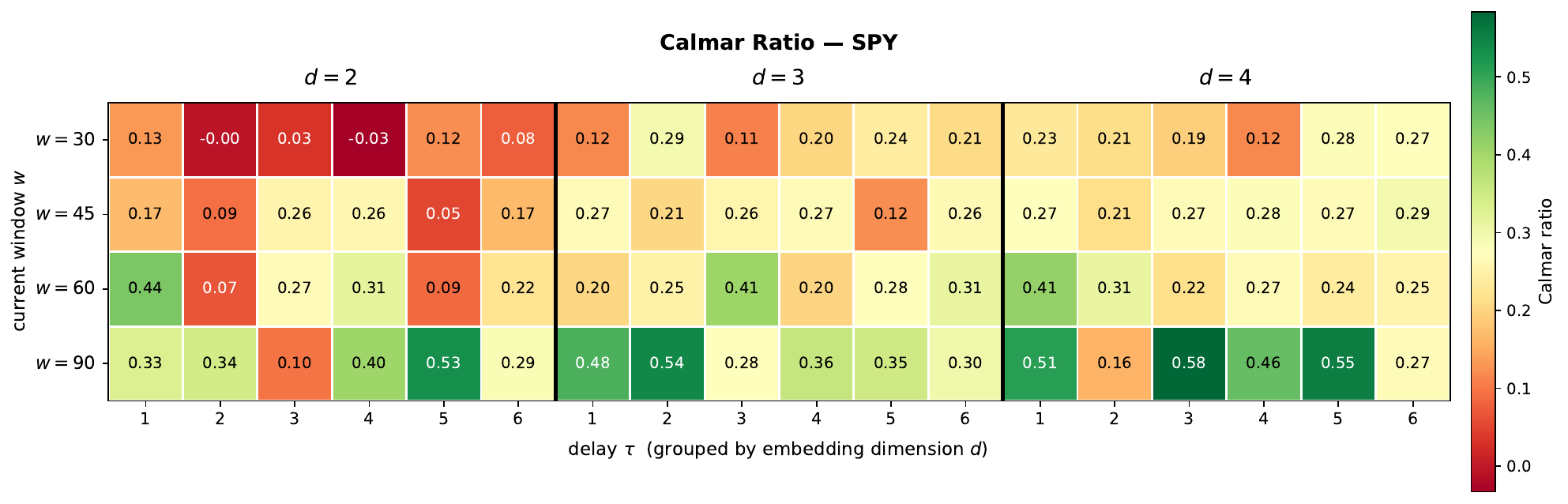}
  \caption{S\&P 500 (SPY).}
\end{subfigure}
\caption{Calmar ratio for every configuration of the Takens-embedding grid
search; layout as in Figure~\ref{fig:grid-heatmap}.}
\label{fig:grid-calmar}
\end{figure}
\FloatBarrier
\end{document}